\documentclass[aps,prl,twocolumn,superscriptaddress]{revtex4-1}
\usepackage{braket}
\usepackage{graphicx}
\usepackage{amsmath}    
\usepackage{amsfonts}
\usepackage[many]{tcolorbox}
\usepackage{subfigure}  
\usepackage{epstopdf}
\def\bf#1{\mathbf{#1}}

\usepackage{lipsum}

\begin{document}

\title{Quantum maximum power transfer theorem}
\author{Cristian L. Cortes}
\affiliation{Birck Nanotechnology Center and Purdue Quantum Center, \\ School of Electrical and Computer Engineering,\\ Purdue University, West Lafayette, IN 47906, U.S.A.}
\author{Zubin Jacob}
\affiliation{Birck Nanotechnology Center and Purdue Quantum Center, \\ School of Electrical and Computer Engineering,\\ Purdue University, West Lafayette, IN 47906, U.S.A.}

\begin{abstract}
We discover the quantum analog of the well-known classical maximum power transfer theorem. Our theoretical framework considers the continuous steady-state problem of coherent energy transfer through an $N$-node bosonic network coupled to an external dissipative load. We present an exact solution for optimal power transfer in the form of the \emph{maximum power transfer theorem} known in the design of electrical circuits. We provide analytical expressions for both the maximum power delivered to the load as well as the energy transfer efficiency which are exact analogs to their classical counterparts. Our results are applicable to both ordered and disordered quantum networks with graph-like structures ranging from nearest-neighbour to all-to-all connectivities. This work points towards universal design principles which adapt ideas of power transfer from the classical domain to the quantum regime for applications in energy-harvesting, wireless power transfer, energy transduction, as well as future applications in quantum power circuit design.
\end{abstract}

\maketitle
\noindent

\section*{Introduction}
Energy transport through a network of ordered or disordered  sites is a universal problem with wide ranging applications in energy transduction, energy harvesting, energy transport through turbid media, as well as information transfer in communication networks \cite{kim2012maximal,popoff2014coherent,shi2012transmission,rebentrost2009environment,engel2007evidence,trautmann2018trapped,trautmann2018trapped,viciani2015observation,viciani2016disorder,ho2017critical,panitchayangkoon2011direct,huber2018anomalous,krikidis2014simultaneous,zhang2013mimo,ng2013wireless,varshney2008transporting,grover2010shannon,desoer1973maximum}. Of particular importance is the development of physical and engineering principles that guide the design of quantum optical networks or nanostructured devices which optimize performance metrics like the total power delivered to a load or the total energy transfer efficiency \cite{caruso2009highly,chin2010noise,olaya2008efficiency,blum2012nanophotonic,newman2018observation,Tal2018,cortes2018fundamental}. 

In classical electrical circuit design, the maximum power transfer theorem states that in order to obtain maximum power transfer to an external load from a source with finite internal resistance, the load must be designed to have an effective resistance that is \emph{equal} to the source resistance. This concept is known as impedance matching and ensures the electrical circuit will deliver maximum  power to the load. This theorem is a textbook example of an engineering principle that guides the design of electrical circuits, transmission lines, and classical wireless communication networks \cite{ng2013wireless,varshney2008transporting,grover2010shannon}. For instance, impedance matching is often used to maximize the transmission of classical information through a communications network because optimal signal delivery requires a signal with maximum strength at the receiving end. In contrast, quantum networks aim to transmit quantum information, for example a state $\ket{\phi}$, from a sender to a receiver with maximum fidelity. It is natural to ask whether similar guiding principles exist for quantum systems, specifically, whether the classical power transfer theorem can be used in the context of quantum networks. 

Here, we present a first step towards the development of a unified framework for power delivery in quantum networks. We introduce an $N$-node coupled bosonic system which is coupled to an external dissipative load, shown in Figure 1, as the quantum analog of a classical network. Physically, the quantum network can be thought of as a system of coupled optical cavities or waveguides or, alternatively, as a system of coupled two-level systems operating in the weak-excitation regime \cite{olaya2008efficiency,wang2018quantum,chin2018coherent}. The load can be thought of as an outcoupling fibre, waveguide, or an external detector. We show that the problem of optimizing the power delivered to the load can be understood through the context of the maximum power transfer theorem resulting in an intuitive impedance matching condition for maximum power delivery. In simple form, we state the quantum maximum power transfer theorem as follows: 

\emph{A load connected to a linear quantum network receives maximum power when the effective load Hamiltonian is conjugately matched to the Thevenin equivalent Hamiltonian of the network.}


As we show below, the Thevenin equivalent Hamiltonian arises from the equivalent single-node representation of the quantum network as seen from the load. It can also be understood as a direct result of the linear bosonic properties of the network, described by Hamiltonian (1). The results we present in this manuscript are valid in the weak and strong coupling regimes and are applicable in both the incoherent and quantum coherent energy transfer regimes. Furthermore, we show that the theorem applies to both ordered or disordered networks with graph-like structures ranging from nearest-neighbor connectivity to all-to-all connectivity. Our model also takes into account dissipation channels which are unavoidable in realistic systems. 

\begin{figure}[t!]
	\includegraphics[width=8cm]{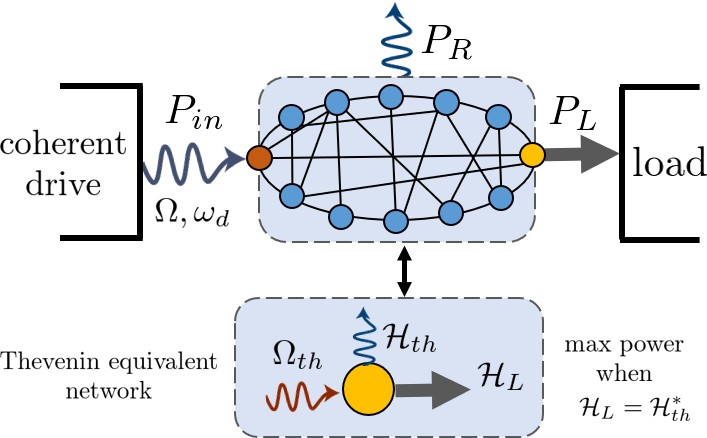}
	\caption{Power transfer model through a dissipative quantum network described by non-Hermitian Hamiltonian $\mathcal{H}_{eff}$ (top) with Thevenin equivalent network (bottom). Maximum power transfer occurs when the non-Hermitian Hamiltonian of the load is conjugately matched with the Thevenin equivalent Hamiltonian, $\mathcal{H}_{th}$. The load may represent an outcoupling fibre, waveguide, or a detector.}
\end{figure}


\vspace{-0.4cm}
\subsection*{Description of dissipative quantum network}

We consider a general model of power transfer through a bosonic quantum network described by the Hamiltonian ($\hbar = 1$),
\begin{equation}
 H = \sum_n \tilde \omega_n\hat{a}_n^\dagger\hat{a}_n + \sum_{n,m} J_{nm}\hat{a}_n^\dagger\hat{a}_m  +  \Omega_1 \hat{a}_1 e^{i\omega_d t} + \text{h.c.} 
\end{equation}
The bosonic excitation is described by operator $\hat{a}_n^\dagger$ ($\hat{a}_n$) which creates (destroys) a boson at site $n$ with frequency $\tilde\omega_n = \omega_n + \delta\omega_n$, satisfying the commutation relation $[a_n,a_m^\dagger] = \delta_{nm}$. Here, $\delta\omega_n$ represents a frequency shift, a so-called bosonic lamb shift, arising from coupling from the $n$th node to the load. The excitation travels through the network with hopping amplitude $J_{nm}$ between the $n$th and $m$th nodes. The first node of the network is driven by a coherent driving field with frequency $\omega_d$ and drive amplitude $\Omega_1$. 


Treating dissipation in the Born-Markov limit, the density operator $\rho$ obeys the Lindblad master equation
\begin{equation}
	\dot \rho = -i[H,\rho] + \mathcal{D}_{R}[\rho] + \mathcal{D}_{L}[\rho].
\end{equation}
The first term on the right represents the coherent evolution of the total system. The second term describes the undesirable (radiative) dissipation of the network. The last term describes dissipation from the $n$th node to an external load. Both terms are explicitly written in terms of the Lindblad superoperators, 
\begin{align}
	\mathcal{D}_{R}[\rho] &=  \sum_{n}\gamma_{n}(\hat{a}_n\rho \hat{a}_n^\dagger - \frac{1}{2}\{\hat{a}_n^\dagger\hat{a}_n, \rho\}) \\
	\mathcal{D}_{L}[\rho] &= \sum_{n}\Gamma_{n}(\hat{a}_n\rho \hat{a}_n^\dagger - \frac{1}{2}\{\hat{a}_n^\dagger\hat{a}_n, \rho\}).
\end{align}
The decay rate $\gamma_n$ represents the undesirable intrinsic decay rate of the $n$th node, while $\Gamma_n$ represents the decay rate from the $n$th node to the load. 
Throughout this manuscript, we work in regimes where $J_{nm},\gamma_{n},\Gamma_n \ll \omega_n$, consistent with the Born-Markov approximation.

\subsection*{Power and efficiency}
The power flow through the system is obtained from the Hamiltonian equation of motion, $\partial_t\!\braket{H} =  \text{tr}( \dot H\rho ) + \text{tr}( H\dot\rho ) = P_{in} - P_{out}$ \cite{kosloff2013quantum,kosloff2014quantum}. Under steady-state conditions, $\partial_t \braket{H} = 0$, the quantum network satisfies the power balance relation, $P_{in} = P_{out}$. The input power $P_{in} = \text{tr}(\dot H\rho )$ represents power coupled into the network and is given by
\begin{align}
	P_{in}  &= i\omega_d[\Omega_1\braket{a_1}e^{i\omega_d t}-\Omega_1^*\braket{a_1^\dagger}e^{-i\omega_d t}].
\end{align}
The output power is divided into two major contributions, $P_{out} = P_{R} + P_L$, corresponding to power dissipated within the network $P_{R} = \text{tr}\left(H\mathcal{D}_{R}[\rho]\right)$ and the power delivered to the load, $P_{L} = \text{tr}\left(H\mathcal{D}_{L}[\rho]\right)$. As shown in the appendix, these equations can be written in the following form,
\begin{align}
    P_R &= \tfrac{1}{2}\sum_{n,m}(\gamma_n + \gamma_m)\left[ \mathcal{\omega}_{nm}\braket{a_n^\dagger a_m} +  \Delta_{nm}\braket{a_n^\dagger}\braket{a_m} \right], \\
    P_L &= \tfrac{1}{2}\sum_{n,m}(\Gamma_n + \Gamma_m)\left[ \mathcal{\omega}_{nm}\braket{a_n^\dagger a_m} +  \Delta_{nm}\braket{a_n^\dagger}\braket{a_m} \right]  
\end{align}
where $\Delta_{nm} = \delta_{nm}\omega_{d} - \omega_{nm}$. In the absence of either dephasing or an incoherent pump, such as a thermal bath, the coherent drive implies the factorization condition, $\braket{a^\dagger_n a_m} = \braket{a^\dagger_n}\braket{a_m}$, is exact in the steady-state regime. In this limit, all relevant observables can be written in terms of the field amplitudes $\braket{a_n}$. Accordingly, the radiated and load power simplify to:
\begin{align}
    P_R &= \omega_d\sum_{n} \gamma_n\braket{a_n^\dagger}\braket{a_n}, \\
    P_L &= \omega_d\sum_{n} \Gamma_n \braket{a_n^\dagger}\braket{a_n}.
\end{align}
Note that the radiated and load power are proportional to the driving frequency $\omega_d$. In the general case of incoherent pumping or dephasing, there will be contributions that are proportional to both $\omega_d$ and $\omega_n$ since $\braket{a^\dagger_n a_m} \neq \braket{a^\dagger_n}\braket{a_m}$. This can be readily confirmed by comparing eqs. (13) and (14) in Appendix B. Finally, we also define the energy transfer efficiency as
\begin{equation}
	\eta = \frac{P_{L}}{P_{in}} = \frac{P_{L}}{P_L + P_{R}}
\end{equation}
representing the percentage of power transferred into the load. 

\subsection*{Thevenin equivalent network}
\noindent
We now show that when a single node (here denoted as the $N$th node) is connected to the load with all other nodes having zero coupling to the load, it is possible to rewrite the equations of motion for the $N$-node network as a \emph{single-node} Thevenin equivalent equation of motion. As we show in the next section, this allows the solution of the optimization problem for maximum power transfer to be understood as an intuitive impedance matching condition. To demonstrate the existence of a Thevenin equivalent network, we first write the equations of motion for the field amplitudes in matrix form (see Appendix B),
\begin{equation}
	i\boldsymbol{\Omega} = (\boldsymbol{\mathcal{\tilde H}} + \boldsymbol{\mathcal{H}}_L)\mathbf{\tilde a},
\end{equation}
where the matrix elements of the non-hermitian matrix $\mathcal{\boldsymbol{\tilde{H}}}$ are $\tilde{\mathcal{H}}_{nm} = i( \delta_{nm}\omega_d-\omega_{nm}) - \delta_{nm}\gamma_{nm}/2$, $\mathbf{\Omega} = (\Omega_1,\cdots,0)^T$, and $\mathbf{a} = (\braket{a_1},\cdots,\braket{a_N})^T$. For notation purposes, we have written $J_{nm}$ as $\omega_{nm}$ when $n\neq m$. Here, $\mathcal{\boldsymbol{{H}}}_L$ is a diagonal matrix representing the non-Hermitian coupling from the $n$th node to the load with diagonal elements ${\mathcal{H}_L}_{nn} = i\delta\omega_{n} - \Gamma_{n}/2$. 
Repetitive back substitution produces the following steady-state equation for $\braket{a_N}$,
\begin{align}
    i\Omega_{th}^{(N)} = \mathcal{\tilde H}_{th}^{(N)}\braket{\tilde a_N} + \mathcal{H}_L\braket{\tilde a_N}
\end{align}
where $\mathcal{H}_{th}^{(N)}$ is interpreted as the Thevenin equivalent energy of the $N$th node. Using the Sherrman-Morrison formula, the explicit expression for the Thevenin equilavent energy is:
\begin{equation}
	\mathcal{\tilde H}_{th}^{(N)} = (\bf{e}_N^T \boldsymbol{\mathcal{\tilde H}}^{-1} \bf{e}_N)^{-1}
\end{equation}
where we also define the Thevenin equivalent Rabi frequency, ${\Omega}_{th}^{(N)}$, as
\begin{equation}
    \Omega_{th}^{(N)} = \frac{\bf{e}_N^T \boldsymbol{\mathcal{H}}^{-1} \bf{\Omega}}{\bf{e}_N^T \boldsymbol{\mathcal{H}}^{-1} \bf{e}_N}.
\end{equation}
Here, $\mathbf{e}_1 = (1, 0, \cdots, 0)^T$ and $\mathbf{e}_N = (0, 0, \cdots, 1)^T$ denote the unit vectors of the first and last nodes respectively. We point out that the non-hermitian Hamiltonian plays an analogous role to complex impedance in electrical circuits. The Thevenin equivalent energy is the quantum network generalization of the Thevenin impedance. 

In summary, we have shown that the quantum network can be re-written as an equivalent single-node network with equivalent driving field (Thevenin Rabi frequency) and equivalent self-energy (Thevenin energy) for the $N$th node. Explicitly, the mean amplitude of the $N$th node is $\braket{\tilde a_N} = (\mathcal{\tilde H}_{th} +i \frac{\Gamma_N}{2})^{-1}\Omega_{th}.$ These results represent the exact solution to the driven-dissipative $N$-node problem for quantum networks with arbitrary couplings and graph-like structures. 


\subsection*{Maximum Power Transfer Theorem}
Following the results from the previous two sections, the power delivered to the load is: $P_L = \omega_d \Gamma_N \braket{a_N^\dagger}\braket{a_N}$. The Thevenin equivalent  representation implies the general solution for an $N$-node network is:
\begin{equation}
    P_L = \omega_d \Gamma_N \frac{|\Omega_{th}|^2}{|\mathcal{\tilde H}_{th} +\mathcal{H}_L|^{2}}.
\end{equation}
The maximum power delivered to the load is then found by optimizing the load's induced frequency shift $\delta\omega_N$ and decay rate $\Gamma_N$, using $\frac{\partial P_L}{\partial \delta\omega_N} = 0$ and $\frac{\partial P_L}{\partial \Gamma_N} = 0$. After a bit of algebra, we find maximum power transfer occurs when
\begin{equation}
	\delta\omega_N = -\delta\tilde{\omega}_{th} \;\;\text{and}
	\;\; \Gamma_N = \Gamma_{th},
\end{equation}
or equally,
\begin{equation}
    \mathcal{H}_L = \mathcal{\tilde{H}}_{th}^*.
    \label{ImpedanceMatching}
\end{equation}
This condition is equivalent to the so-called \emph{conjugate impedance matching} condition that is known in circuit theory. Note that this impedance matching condition depends only on the non-Hermitian Hamiltonian parameters of the quantum network and is not dependent on the driving field Rabi frequency, $\Omega_1$. When this condition is satisfied, the maximum power delivered to the load is 
\begin{equation}
	P_{L,max} = \hbar\omega_d\frac{|\Omega_{th}|^2}{\Gamma_{th}}.
	\label{MaxPower}
\end{equation}
Equations (\ref{ImpedanceMatching}) and (\ref{MaxPower}) form the main results of the paper. From this expression, it is clear that quantum networks with a large Thevenin Rabi frequency $\Omega_{th}$ and small Thevenin dissipative term, $\Gamma_{th}$ are ideal for enabling large power transfer rates.

\subsection*{Energy Transfer Efficiency}
While the conjugate impedance matching condition (\ref{ImpedanceMatching}) ensures maximum power transfer, it does not guarantee maximum energy transfer efficiency. This result is exactly analogous to the classical case in electrical circuit design. Here, we present an exact form for the energy transfer efficiency given by, 
\begin{equation}
    \eta = \frac{ \Gamma_N }{\mathcal{F}\gamma_1 + \gamma_n + \Gamma_N},
\end{equation}
where $\mathcal{F} = |\mathcal{H}_{22} + \mathcal{H}_L|^2/|H_{12}|^2$. In general we find that the energy transfer efficiency cannot be written in a form that can be understood from the single-node Thevenin equivalent network representation. In the limit of large effective coupling, $H_{12}$, between the first and last nodes, the energy transfer efficiency is limited primarily by the radiative loss of the $N$th node, $\gamma_N$m as one might expect. Perhaps more importantly, it is also possible to prove that the energy transfer efficiency, when the impedance matching condition is satisfied, will always be less than or equal to fifty percent. This demonstrates the existence of a fundamental trade-off between the maximum power that can be delivered to the load and the energy transfer efficiency.

\begin{figure*}[ht!]
	\includegraphics[width=18cm]{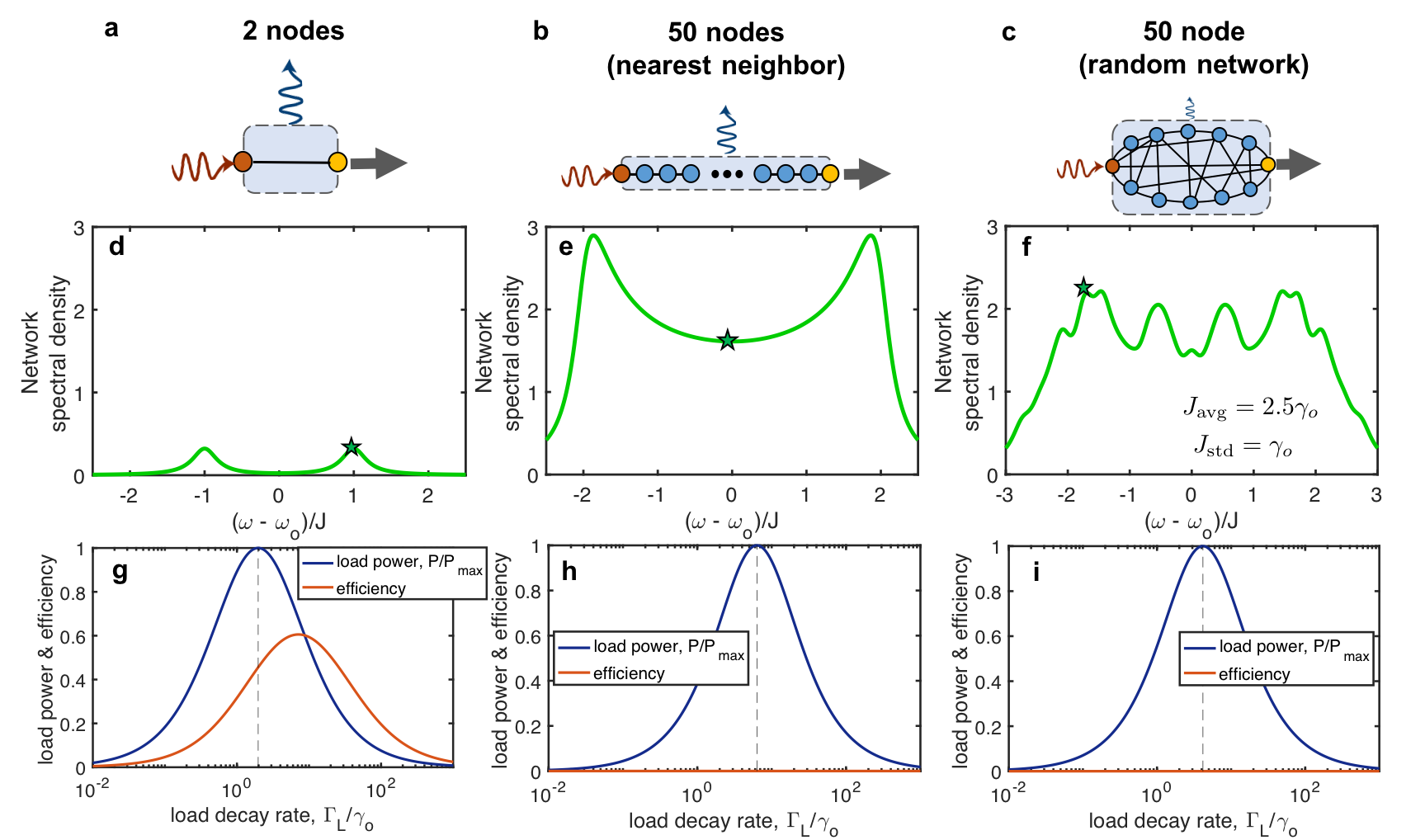}
    \caption{Demonstration of quantum maximum power transfer theorem for three distinct networks: (a) 2-node network, (b) 50-node nearest neighbor network, (c) 50-node network with random all-to-all connectivity. (d)-(f) Network spectral density as a function of the input drive frequency $\omega$. All nodes are assumed to have resonant frequency $\omega_o$. The network spectral density shows the relative importance of different spectral modes within the network, and is closely related to the transmission amplitude used in input-output theory.  (g) - (i) Numerical simulations of the load power and efficiency as a function of load decay rate. Each network is driven by a coherent field with driving frequency ($\omega_d = \{ \})$ denoted by the green stars in (d)-(f). Note the numerical simulations show excellent agreement with the quantum impedance matching condition (15). }
\end{figure*} 

\section*{Discussion}

In figure 2, we present full numerical simulations for three distinct quantum networks including: (a) a simple 2-node network, (b) a 50-node 1-dimensional chain with nearest-neighbor coupling, and (c) a 50-node network with random all-to-all connectivity. The hopping parameters $J_{nm}$ of the 50-node random network are sampled from a normal distribution with mean $J_{avg} = \tfrac{5}{2}\gamma_o$ and standard deviation $J_{std} = \gamma_o$. To characterize different quantum networks, we use the network spectral density, $	S(\omega_d) = \text{Im}[\text{tr}(\omega_d - \mathbf{\mathcal{H}})^{-1}]$, which is closely related to the transmission amplitude which is well-known in input-output theory \cite{plankensteiner2017cavity}. 

The network spectral density in figures 2 (e)-(f) show the relative magnitudes of the network's eigenmodes. For the two-node network shown in Figure 2 (a), there exists two dominant modes known as the symmetric and anti-symmetric modes with resonant frequencies $\omega_o + J$ and $\omega_o - J$ respectively (see Figure 2-d). Furthermore, a driving field with frequency $\omega_d$ can be used to couple to a particular eigenstate of the network. In figures 2 (g)-(i), we simulate driving the quantum network with different frequencies for each network, highlighted by the green stars in figures 2 (d)-(f). By using the quantum impedance matching condition (17) as well as the Thevenin equivalent self-energy (13), we are able to calculate the optimal load decay rate where maximum power transfer occurs. The optimal load decay rates predicted by equation (13) are given by the vertical dashed lines in Figure 2 (g)-(i). As expected, the exact analytical expressions match exactly with full numerical simulations.

\section*{Conclusion}
In summary, we have presented the quantum analog of the maximum power transfer theorem for an $N$-node bosonic network. In recent years, research into energy transfer through complex networks, such as those arising in photosynthesis, have led to the development of reconfigurable and programmable energy transport simulators \cite{trautmann2018trapped,potovcnik2018studying,viciani2016disorder,viciani2015observation,chin2018coherent,wang2018quantum,gorman2018engineering}. This work can therefore be tested immediately with reprogrammable photonic networks or other quantum simulators of bosonic systems. Finally, while this work provides a simple rule of thumb for linear networks, the role of nonlinearity as seen in Bose-Hubbard Hamiltonian models \cite{kuhner1998phases,goral2002quantum,kollath2007quench,buonsante2007ground} requires additional considerations which should be studied carefully in the near future. 

\section*{Acknowledgements}

This work was supported by the Defense Advanced Research Projects Agency (grant number N66001-17-1-4048).


\bibliography{QMPTTrefs.bib}

\end{document}